\documentclass[twocolumn]{aastex62}
\usepackage{rotating}

\newcommand{\kmps}{\,km\,s$^{\rm -1}$}

\newcommand{\pmode}{{\it p}-mode}

\newcommand{\Fe}{Fe {\sc i} }

\shorttitle{Internal Excitation of Sunspot Oscillations}
\shortauthors{Cho et al.}

\begin{document}

\title{The Observational Evidence for the Internal Excitation of Sunspot Oscillations Inferred from the  \Fe 5435 \AA\ Line}

\author[0000-0001-7460-725X]{Kyuhyoun Cho}
\email{chokh@astro.snu.ac.kr}
\affil{Astronomy Program, Department of Physics and Astronomy, Seoul National University, Seoul 151-747, Republic of Korea}

\author[0000-0002-7073-868X]{Jongchul Chae}
\affil{Astronomy Program, Department of Physics and Astronomy, Seoul National University, Seoul 151-747, Republic of Korea}

\author[0000-0002-7358-9827]{Eun-kyung Lim}
\affil{Korea Astronomy and Space Science Institute, Daejeon 34055, Republic of Korea}

\author[0000-0001-5455-2546]{Heesu Yang}
\affil{Korea Astronomy and Space Science Institute, Daejeon 34055, Republic of Korea}

\correspondingauthor{Jongchul Chae}
\email{jcchae@snu.ac.kr}



\begin{abstract}
The umbral oscillations of velocity are commonly observed in the chromosphere of a sunspot. Their sources are considered to be either the external \pmode\ driving or the internal excitation by magnetoconvection. Even though the possibility of the \pmode\ driving has been often considered, the internal excitation has been rarely investigated.  We report  the identification of the oscillation patterns that may be closely related to the events of internal excitation from the observations of velocity oscillations in the temperature minimum region of two sunspots. The velocities were determined from the  spectral data of the \Fe 5435 \AA~line, a magnetically insensitive line,  taken with the Fast Imaging Solar Spectrograph of the 1.6 m Goode Solar Telescope at the Big Bear Solar Observatory. As a result, we identified  4 oscillation patterns of  $2.0 \times 10^3$ km coherent size that were clearly identified for about 7.9 minutes with oscillation amplitude of $9.3 \times 10^{-2}$ \kmps.   The power of the oscillations in these centers was concentrated in the 3 minute band. All the oscillation centers were located above the umbral dots undergoing noticeable morphological and dynamical changes that may be regarded as an observable signature of small-scale magnetoconvection inside the umbrae. Our results support the notion that  magnetoconvection associated with umbral dots inside sunspots can drive the 3 minute umbral oscillations.
\end{abstract}

\keywords{Sun: oscillations --- sunspots --- Sun: photosphere}


\section{Introduction} \label{sec:intro}

Velocity oscillations are commonly observed in sunspot umbrae at different atmospheric levels. They are generally interpreted as slow magneto-acoustic waves propagating upwards from the lower layer \citep{Centeno2006, Felipe2010}. At the photospheric level their power spectra are similar to that of the quiet Sun in shape, but suppressed in magnitude \citep{Howard1968, Abdelatif1986}. Most of their power is concentrated on the 5 minute period, and  small power exists in the 3 minute band. With height, the 5 minute oscillations become weaker due to the acoustic cutoff, and hence the 3 minute oscillations become dominant in the chromosphere \citep{Bogdan2006}. This is an outcome of linear propagation where the sunspot oscillations of different frequencies propagate independently with no power exchange between the 5 and 3 minute oscillations \citep{Lites1986}. The surviving 3 minute oscillations nonlinearly develop into shocks at a chromospheric height \citep{Chae2014} and propagate upwards to the chromosphere-corona transition region or above \citep{Tian2014, Jess2012}.

It is very likely that the 3 minute  umbral oscillations are driven either in the photosphere or below the surface because they are observed to propagate upward in the chromosphere. There are two candidates for the driving. The first is the external \pmode\ driving. Even though  the \pmode\  has a  peak of power around the 5 minute period,  its higher frequency tail may contain some power at the 3 minute period.  Observational results have been reported that are consistent with this driving, which include the similarity of the frequency spectrum and oscillating pattern with the quiet Sun \citep{Zhao2013, KrishnaPrasad2015}, the \pmode\ absorption coefficient \citep{Braun1995}, the inward traveling wave power at the umbra-penumbra boundary \citep{Penn1993}, and the direct detection of peculiar velocity pattern \citep{Beck2010}.

The other candidate is the internal excitation, probably by the small-scale magnetoconvection inside umbrae. Despite strong magnetic field, it has been theoretically expected that small-scale magnetoconvections can take place in the photosphere of an umbra \citep{Schussler2006}. In fact, convective motions are expected to occur more easily beneath the photospheric layer because the high $\beta$ plasma condition for convection to occur is better fulfilled in the interior. Theoretical studies anticipated that acoustic waves can be generated in the convective environment \citep{Lee1993, Moore1973} and the acoustic power will be the most observable near the cut-off frequency in the chromosphere \citep{Chae2015}.

Recently, upward and downward motions near the umbral dots were reported \citep{Watanabe2012, Ortiz2010}, which may be regarded as the observational evidence for magnetoconvection inside umbrae. Nevertheless, for long time there have been few observational reports supporting the direct connection between the umbral oscillations and the magnetoconvection. Only very recently such evidence was reported by \citet{Chae2017}. They found the enhancement of the 3 minute oscillation power near the light bridge and umbral dots of a sunspot. Since the light bridge and umbral dots are commonly regarded as the observable features of magnetoconvection, this observation can be considered to be supporting the internal excitation by magnetoconvection.

It may be reasonable to suppose that the excitation of waves, being irrespective of whether it is internal or external,  takes place in the form of excitation events  localized in space and time.  In fact, such excitation events of  the \pmode s were reported to  occur in the quiet regions of the Sun \citep{Goode1992, Rimmele1995}.  In the case of the internal excitation,  the source should be located inside sunspot umbrae, either in the photosphere or in the interior below it.  Each event will drive oscillations and cause them to propagate vertically and/or horizontally. As a result, in a horizontal plane near the photosphere the oscillations will either appear as a stationary pattern in a local area (if the propagation is fully vertical) or propagate outward from a center (if the propagation is not fully vertical) for a finite duration of time. In either case, one can identify the oscillation center, which is supposedly above the excitation source. This kind of pattern of oscillations, if identified inside sunspots, can be considered as the observational  evidence of the internal excitation.

In this paper, we report the detection of such oscillation patterns  by analyzing the imaging-spectral data of the \Fe 5435 \AA\ line  taken from two sunspots.  The \Fe 5435 \AA\ line is a strong  and magnetically insensitive absorption line that is formed in the temperature minimum region,  so it is very suited for the investigation of the low atmospheric level behavior of the 3 minute oscillations inside sunspots.

\section{Data and Analysis} \label{sec:anal}

The \Fe 5435 \AA\ spectral data were taken with  the Fast Imaging Solar Spectrograph (FISS; \citealt{Chae2013}) of the Goode Solar Telescope (GST) at the Big Bear Solar Observatory (BBSO). We observed velocity oscillations in two sunspots  on 16 June 2015 and 15 June 2017, respectively.
The target of the 2015 observation was a leading sunspot of AR 12367 (-70\arcsec, -355\arcsec) which was in the fully developed middle stage. The 16\arcsec\ $\times$ 40\arcsec\ field of view was observed with a time cadence of 16 s for 38 minutes from 18:36 - 19:14 UT. \citet{Chae2017} employed this data for investigating umbral oscillation study. In the 2017 observation case, we observed a leading sunspot of AR 12663 (25\arcsec, 205\arcsec). The sunspot was at its early stage, being less than one day after its formation. The 13\arcsec\ $\times$ 40\arcsec\ field of view was observed with a time cadence of 14 s for 50 minutes from 20:21 - 21:10 UT. Both the sunspots were located near the Solar disk center, so that the projection effect is negligible. The seeing condition was good and stable for both the observations. As a result, two sets of four-dimensional data $I(\lambda, x, y, t)$ were obtained from the FISS observations.

Basic data reduction was done following the standard procedure of \citet{Chae2013}. All the FISS data were spatially aligned with the \textit{Helioseismic and Magnetic Imager} (HMI, \citealt{Schou2012}) continuum image. We obtained the time series of the line-of-sight velocity at every pixel in the field-of-view by performing the Gaussian fitting to the \Fe line core.  Next,  we applied a 1-to-4 minute-period bandpass filtering to this time series data  using  the package of wavelet analysis  provided by \cite{Torrence1998},   to minimize of the effects of   noise and the 5 minute oscillations. 

\begin{figure*}
\epsscale{1}
\centerline{\includegraphics[width=1 \textwidth,clip=]{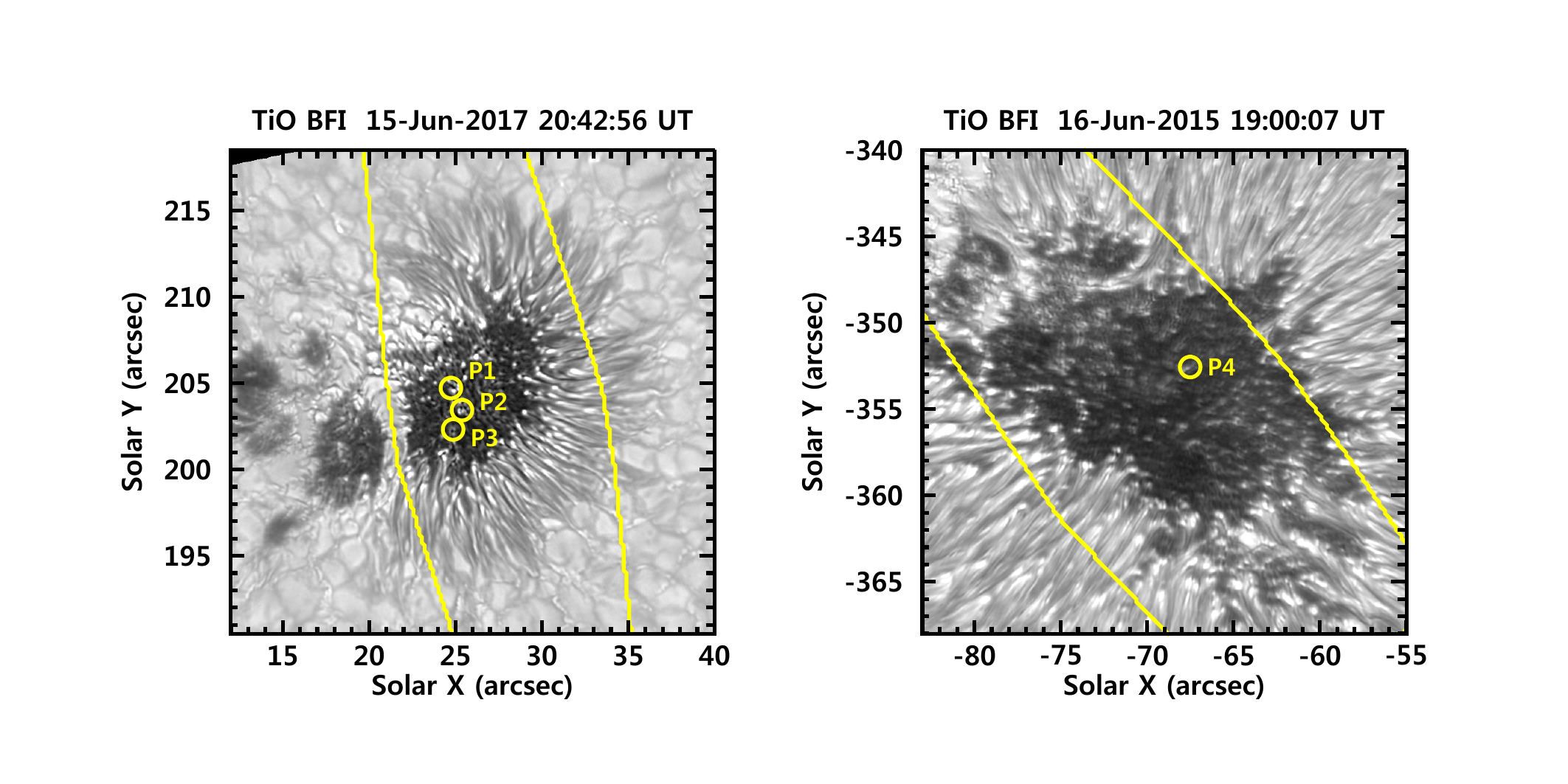}}
\caption{TiO 7057 \AA\ broadband filter images of the two sunspots. The yellow polygons represent the FISS field of view. The locations of the oscillation centers are marked by the yellow circles annotated as P1 to P5, respectively. Each oscillation center number represents the event number.}
\label{fig1}
\end{figure*}

Together with the FISS data, we use the speckle-reconstructed TiO 7057 \AA\ broadband filter images \citep{Cao2010} to analyze fine structures and changes in the photospheric level of the umbrae. The TiO filter has a 10 \AA\ bandwidth and a pack of  70 TiO images was taken every 20 s for the 2015 observation and every 15 s for the 2017 observation with an exposure time of 0.7 ms. The obtained TiO data went through the basic data reduction process and then the code of the Kiepenheuer-Institute Speckle Interferometry Package \citep{Woger2008} was applied to make a speckle-reconstructed image from each pack. As a result, we have reconstructed the diffraction-limited data with spatial sampling of 0.017\arcsec. All TiO images are aligned with the reference HMI continuum images for each observation and are then compared with the FISS data for alignment.  

We identify the oscillation patterns  reflecting the excitation events  from the animation of the velocity oscillations in each sunspot (the online animated Figure \ref{fig2-0}).  Supposing that each excitation event occurs below the image plane, that is the formation surface of the \Fe line,  we expect that  there exists a point  in the image plane just above the location of the event.  This point then  becomes an apparent origin like the epicenter of earthquake.  We will refer to this point as the oscillation center. The point is that the velocity oscillation at the center should not be resulting from an apparent horizontal propagation from somewhere else, but should be starting from the point itself. It is very likely that the oscillation center is the center of an apparent outward propagating pattern (ripple), if the waves are produced by the internal excitation below the surface. We trace the propagation of the identified oscillation patterns. We manually choose a number of points located at the edge of the oscillation pattern, then obtain the extent of the oscillation pattern by ellipse fitting. The development of the oscillation pattern can be confirmed in merged image of the ellipse fitting results. The oscillation centers are defined by the peak position of the Doppler velocity inside the first oscillation pattern for each event. The uncertainty of their positions are expected to be less than 0.5\arcsec.  

Some ambiguity in the identification arose because the oscillation pattern at a time was often a superposition of more than one components, driven either internally or externally. Therefore there is a possibility that we have missed some oscillation patterns induced from the internal excitations. For clear identification of the internal excitations, we require the oscillations at the center to last for longer than two periods.

After we identify the oscillation pattern, we determine the start and end of the time interval while the oscillation center can be identified from the velocity animation. We regard the length of this time interval as the duration of the oscillation pattern. Since this determination is based on visual inspection, uncertainties may arise from the ambiguity, which we estimate at about 1 minute for each of the start and end times, and 1.5 minute for the duration. 

We also determine the size of the coherent oscillation pattern around the center by analyzing the two-dimensional spatial autocorrelation of velocity at the time of peak development of the Doppler velocity. The FWHM of the autocorrelation is adopted as the size of the pattern. We think this size is more relevant to the apparent horizontal wavelength than the source size. The horizontal wavelength may be about twice the coherent size.

\section{Result} \label{sec:result}

\begin{figure*}
\epsscale{1}
\centering
\includegraphics[width=1 \textwidth,  clip=]{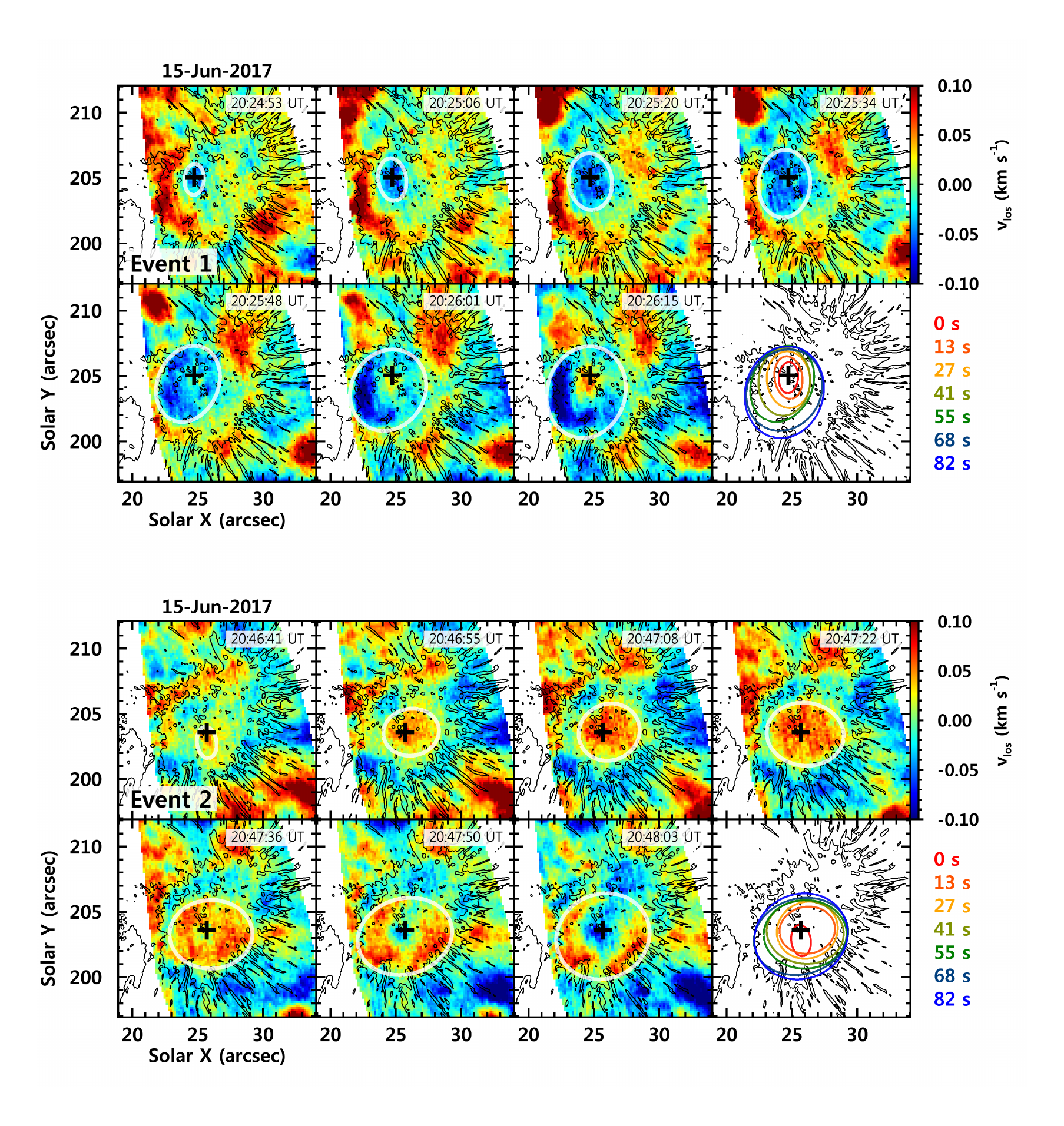}
\caption{Time series of velocity map showing examples of identified oscillation patterns. Upper panel (lower panel)  represents one of the detected oscillation pattern of event 1 (event 2) and their evolution.  Each observation time is shown in upper right part of the every image. The black contours indicate umbral-penumbral boundary and umbral dots extracted from the TiO images for all other figures.  White ellipses indicate the fitting results of the outward propagating oscillation patterns. The black crosses represent the oscillation center. The last image of each panel represents the merged image of the ellipse fitting. The color of the ellipses indicates time after the emergence of the detected oscillation pattern. A full animated version of the Doppler velocities is available online. The animated version shows all four events, tracking them individually for 7-8 minutes each.}
\label{fig2-0}
\end{figure*}

\begin{figure*}
\epsscale{1}
\centering
\includegraphics[width=1 \textwidth,  clip=]{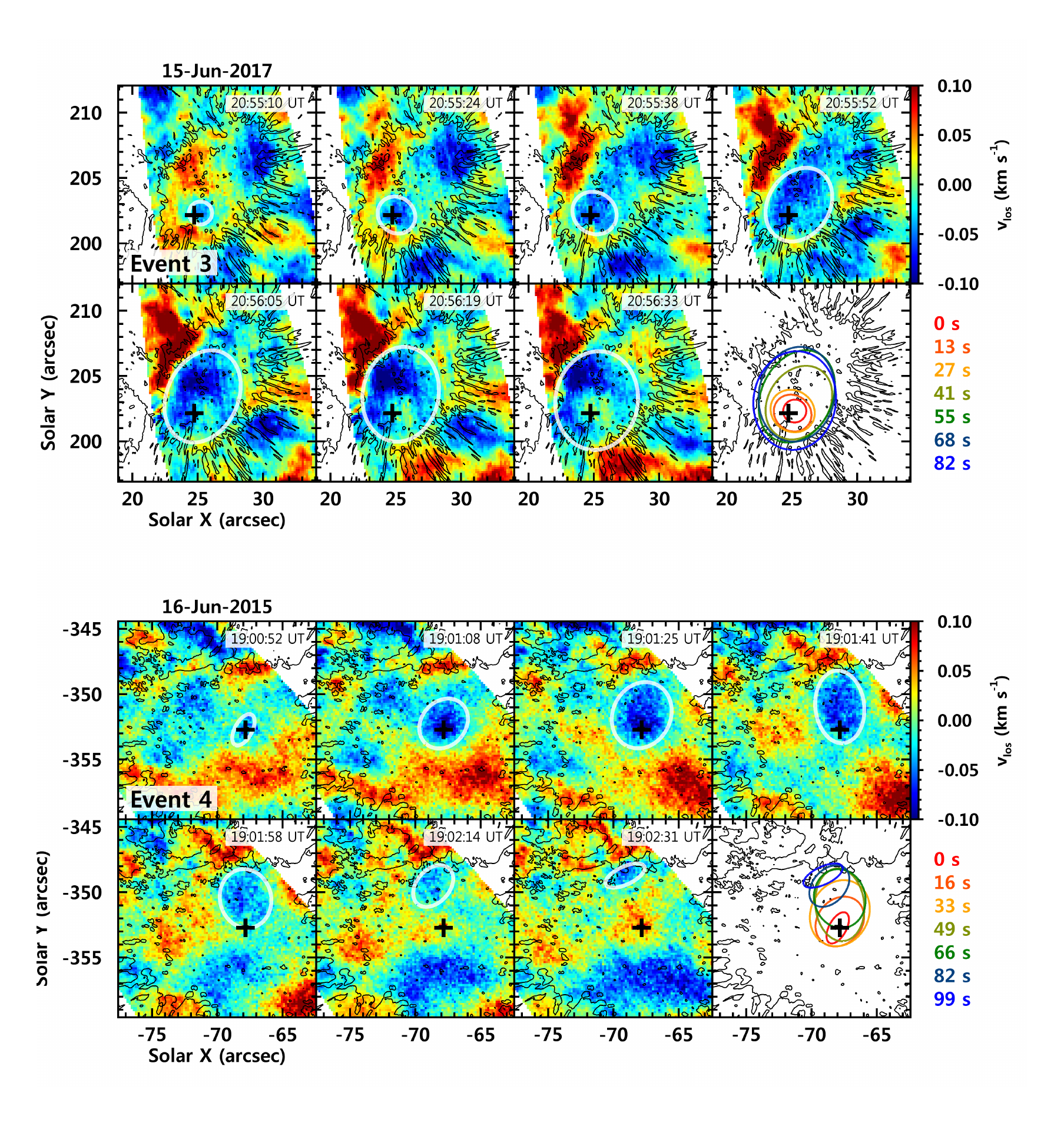}
\caption{Same as the Figure \ref{fig2-0}, but for the event 3 and 4. An animated version of these Events is available in Figure~\ref{fig2-0}.}
\label{fig2-1}
\end{figure*}

We have identified a number of oscillation patterns which may be attributed to four excitation events.  The oscillation centers were marked by P1, P2, P3, and P4 in the photospheric images of the two observed sunspots  of Figure \ref{fig1}, and  by the ellipses on the time series of velocity map in Figure \ref{fig2-0} and \ref{fig2-1}.

Figure \ref{fig2-0} and \ref{fig2-1} show examples of the identified oscillation pattern for each event. From the 7 sequence of images, we can see the development process of the oscillation pattern. For example, in the case of the event 1, the oscillation pattern starts to emerge near the oscillation center within confined area (first image). As time passed, the blue or redshift at the oscillation center increase (second and third images), then the oscillation pattern propagates outward from their center and diminishes (after the fourth image). Similarly, the opposite sign of the oscillation pattern emerges at the same position, the oscillation center (sixth and seventh image) and undergoes the same sequence with the previous oscillation pattern in the same manner.  Other events also show similar process with the event 1.

The identification of the oscillation center was greatly aided by the clear presence of the  outwardly propagating  patterns (ripples) of velocity in the case of the three patterns in the 2017 sunspot, as can be seen from  Figure \ref{fig2-0} and upper panel of Figure \ref{fig2-1}, and the associated animation. In events 1, 2, 3, the ripples can be identified without much difficulty so that it is easy to determine the oscillation centers. The associated animation will help the readers to identify the ripples and the oscillations centers in these events. In event 4 (lower panel of Figure \ref{fig2-1}), the ripple is not obvious, probably due to the poorer seeing condition of observations. Nevertheless, a careful examination of the figure and the animation suggests the existence of such a pattern in this event as well, even though it appears spread and deviates from isotropic propagation.

\begin{figure*}
\epsscale{1}
\centerline{\includegraphics[width=1 \textwidth,clip=]{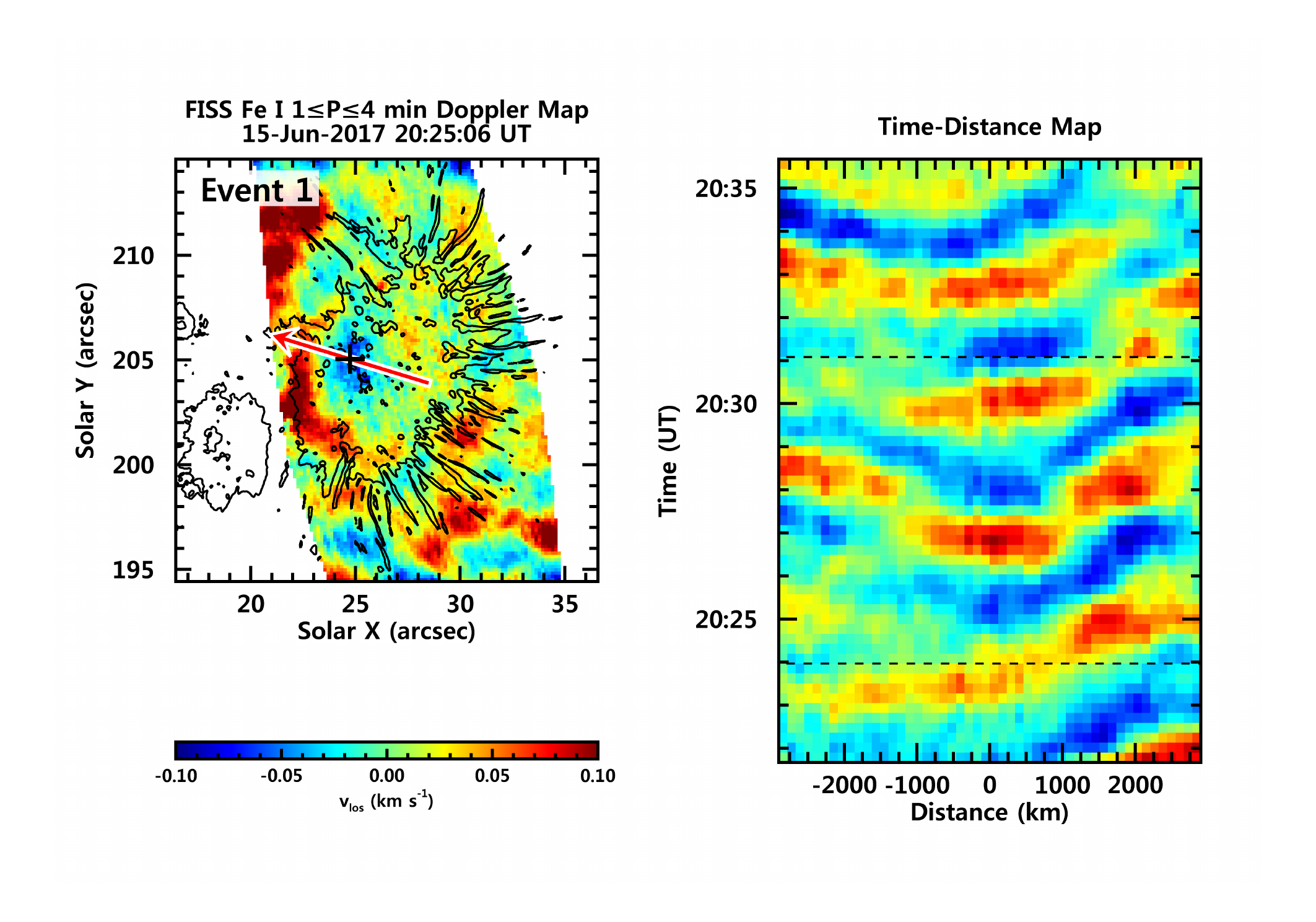}}
\caption{Left: Velocity map for the event 1 at the peak development time. The red arrow indicate the radial direction and the slit position  used for the construction of  the  time-distance map. The black cross represents the oscillation center. Right: Time-distance map of the velocity. The time between black dashed lines indicates the duration of the oscillation patterns. The distance is measured from the oscillation center in the arrow direction.}
\label{fig3}
\end{figure*}

Figure \ref{fig3} presents an example of the time-distance map of velocity for the event 1.  The time-distance map was constructed along the slit that is set to be along the line connecting the umbral center (26.5\arcsec, 204.5\arcsec) and the oscillation center. We identify  $V$-shaped stripes or chevrons in the maps during the time interval of about 7 minutes between 20:24 and 20:31 UT, which was determined for the animation of velocity as the duration of the oscillation patterns. Moreover,  the vertexes are located in the oscillation center. These chevrons are closely related to the ripples mentioned above and a clear observational indication that  the velocity oscillations propagate outward in the two opposite directions from the oscillation center in the image plane. Even though the blue-red patterns are exist before and after the duration, they appear to originate from other region, not the oscillation center. We have examined the time-distance maps constructed along the slit oriented in the perpendicular direction and found similar chevrons in these maps. So we conclude that the orientation of the slit is not critical in this investigation unless the propagation is highly anisotropic.

Other events detected in the 2017 observational data show similar characteristics in their time-distance maps. The event 4 observed in 2015 shows slightly different, because it did not show circular propagating ripples. At the early stage, however, it also exhibits independent oscillation patterns in the confined area near the oscillation center, which is not connected with those in distant region. Then they show unidirectional propagation. (see lower panel of Figure \ref{fig2-1})

We have determined  the observable parameters of the oscillation patterns (Table \ref{tab2-1}).  First,  the amplitude of velocity oscillation at the oscillation centers  was determined to range from  $7.6 \times 10^{-2}$ \kmps\ to $1.2 \times 10^{-1}$ \kmps\ with the mean value of $9.3 \times 10^{-2}$ \kmps. Second, the coherent size of each oscillation patterns at the peak development time ranged from $1.6 \times 10^3$ km to $2.8 \times 10^3$ km with the mean of $2.0 \times 10^3$ km.  Third, the duration of each oscillation pattern ranged from 6.5 minutes to 9.6 minutes with the mean of 7.9 minutes.  Finally,  we determined the speed of the apparent propagation of the oscillations in the image plane from the changes of the ellipse fitting along the fastest expanding direction.  The speed ranged from 13 \kmps\ to 17 \kmps\ with the mean of 15 \kmps.

\begin{deluxetable*}{ccccc}
\tablecaption{Observable parameters of the oscillation patterns
\label{tab2-1}}
\tablecolumns{5}
\tablenum{1}
\tablewidth{0pt}
\tablehead{
\colhead{Event number} & \colhead{Amplitude (\kmps)} & \colhead{Coherent size (km)} & \colhead{Duration (min)} & \colhead{Apparent speed (\kmps)}}
\startdata
 1 & $7.6 \times 10^{-2}$ & $1.8 \times 10^3$ & 7.2 & 15 \\
 2 & $8.6 \times 10^{-2}$ & $1.8 \times 10^3$ & 6.5 & 13 \\
 3 & $8.6 \times 10^{-2}$ & $1.6 \times 10^3$ & 8.2 & 17 \\
 4 & $1.2 \times 10^{-1}$ & $2.8 \times 10^3$ & 9.6 & 15 \\
\hline
Mean & $9.3 \times 10^{-2}$ & $2.0\times 10^3$ & 7.9 & 15 \\
\enddata
\end{deluxetable*}

\begin{figure*}
\epsscale{1}
\centerline{\includegraphics[width=0.9 \textwidth,clip=]{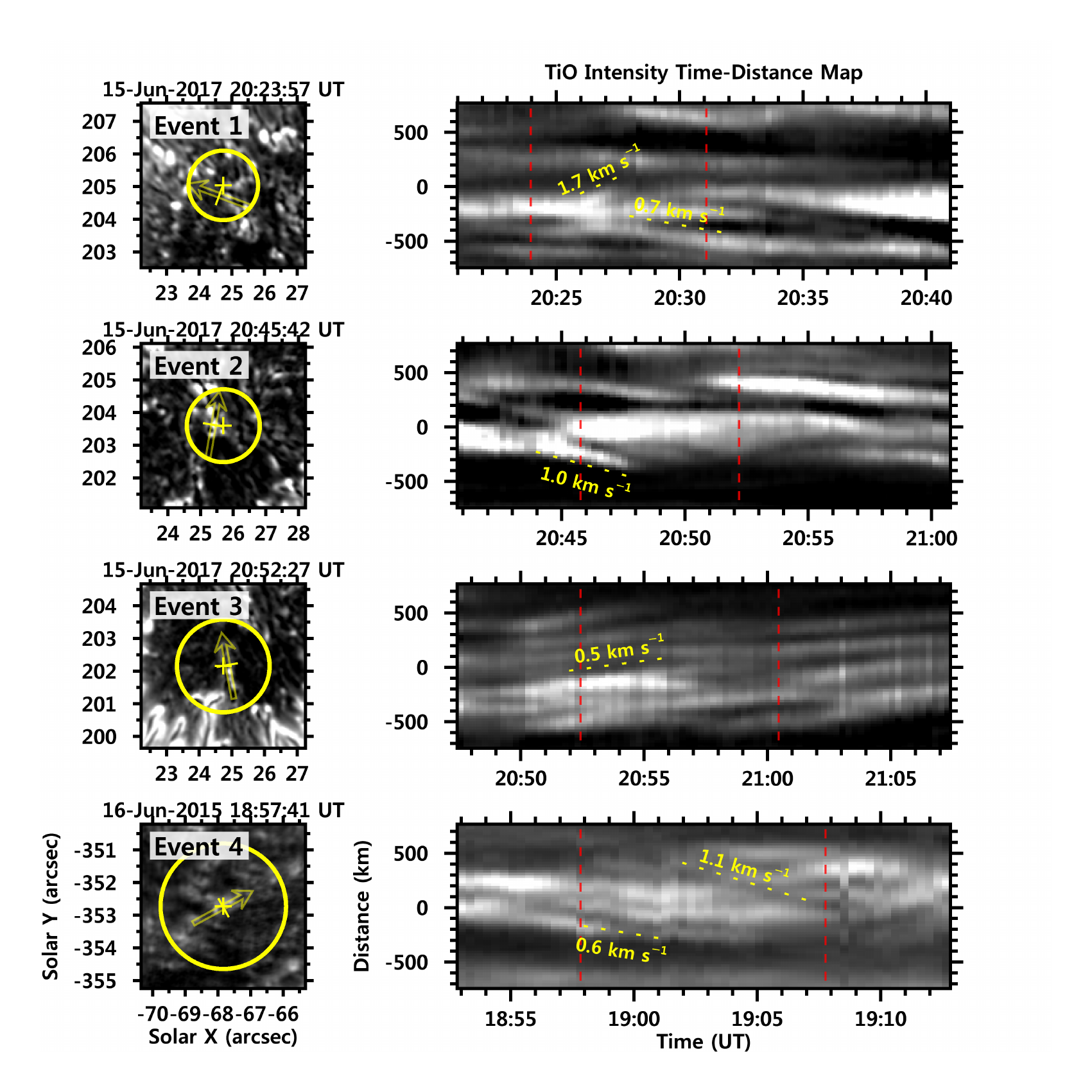}}
\caption{Left: The TiO intensity image at the beginning of the oscillation patterns. The yellow circles and the yellow crosses represent the coherent size of the oscillation patterns and the oscillation centers, respectively. The yellow arrows indicate the slit positions of each time-distance map. The perpendicular tick in the middle of each arrow represents the zero distance which is nearest point from each oscillation center. Right: TiO intensity time-distance map. The time between red dashed lines indicates the duration of the oscillation patterns. The yellow dashed lines indicate the noticeable moving features.  Each row represents different event. An animation of the TiO intensity map for all four events is available online. The animated version shows all four events, tracking them individually for 10-13 minutes each.}
\label{fig4}
\end{figure*}

We found that all the oscillation centers were spatially and temporally associated with  umbral dots.  The three centers (P1, P2, P4)  were near central umbral dots  and  the other (P3) were very close to peripheral umbral dots (see Figure \ref{fig1}).  All the umbral dots were under noticeable complex changes in brightness, shape,  position and so on.

The time-distance maps of photospheric brightness in Figure \ref{fig4} and the associated animation illustrate such complex changes. Each time-distance map was constructed along the slit that was put in the image plane to the direction where the umbral dots moves. The slit positions deviate from the oscillation centers slightly,  by less than 0.34\arcsec\ (250 km) which is comparable to the spatial resolution of the  FISS. These deviations, we think, are not observationally significant.

Near the oscillation center related to the event 1, we observe  umbral dots  brighten, darken, collide, break, move and disappear during the interval between 20:24 and 20:31 UT when the oscillation pattern was clearly identified. At 20:24 UT, we find a bright umbral dot of about 180 km size beneath the oscillation center. Its brightness was increasing with time. In addition to the chief umbral dot,  there existed a smaller and fainter umbral dot 150 km away from this  bright dot at an inner position of the sunspot. The brightness of this dot also gradually increased with time.  From 20:26 UT to 20:27 UT, the two dots were  in contact with each other so they look like a single umbral dot with internal structures.  During this time,  we also  find another dot-like structure emanating from the chief dot outward at a speed of about 1.7 \kmps.  After 20:28 UT,  this structure  developed into another dot that is clearly identifiable.  The chief dot underwent gradual decrease in brightness, and it eventually disappeared. Meanwhile, the  small dot  at the inner position moved away from the bright dot further  inward at a speed of about 0.7 \kmps, and was fully separated again from the chief dot.  

Note that  similar processes occurred in the other oscillation centers, such as brightening (Event 2), breaking (Event 3), and complex moving (Event 4). Particularly the fast motion of the umbral dots happened around all the centers with the mean speed of about 0.9 \kmps, which is much faster than  the typical speed of umbral dots, 0.4 \kmps\ \citep{Riethmuller2008}.  The horizontal distance of umbral dot's migration is typically less than 720 km which may be regarded as the horizontal size of the source. This is smaller than the coherent size of the oscillation patterns.

\begin{figure*}
\epsscale{1}
\centerline{\includegraphics[width=1 \textwidth,clip=]{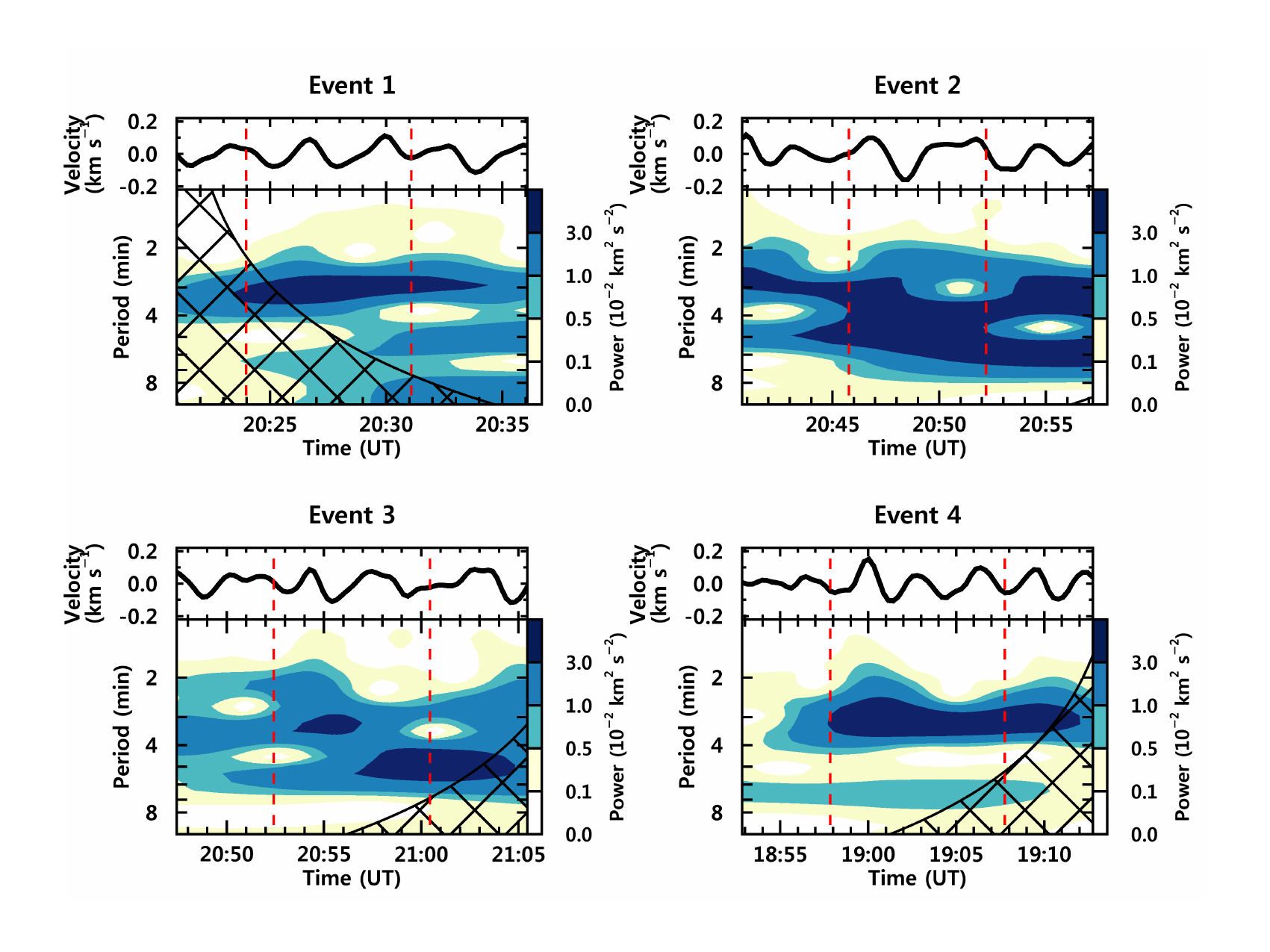}}
\caption{1-to-16 minute-period bandpass filtered FISS \Fe velocity oscillation data at the oscillation center, and its wavelet power spectrum. Each panel represents the different event. The time between red dashed lines indicates the duration of the oscillation patterns. }
\label{fig5}
\end{figure*}

Figure \ref{fig5} presents the time series of 1-to-16 minute-period bandpass filtered FISS \Fe velocity at each center and its wavelet power spectrum. Note that the velocity data in the figure include not only 3-minute (2-to-4 minute) oscillations, but also longer period oscillations. It can be seen from the wavelet spectra that the internally excited oscillations have strong power in the 3-minute band. This is clear from the calculated ratios of the 3-minute power to the total(1-to-16 minute) power presented in Table \ref{tab2-2}, which indicates that  the 3-minute oscillation power is more than half of the total power.


\begin{deluxetable}{ccc}
\tablecaption{Ratios of 3-minute (2-to-4 minute) oscillation power to the total (1-to-16 minute) power temporally averaged over the duration of each event \label{tab2-2}}
\tablecolumns{2}
\tablenum{2}
\tablewidth{50pt}
\tablehead{
\colhead{Event number} & \colhead{Power ratio}}
\startdata
 1 & 0.76 \\
 2 & 0.52 \\
 3 & 0.56 \\
 4 & 0.86 \\
\hline
Mean & 0.67 \\
\enddata
\end{deluxetable}

\section{Discussion} \label{sec:diss}

  We have identified 4 oscillation patterns inside the sunspot umbrae at the temperature minimum level that may be closely related to the events of internal excitation. The oscillation patterns are characterized by the velocity amplitude of about $9.3 \times 10^{-2}$ \kmps, the size of about $2.0 \times 10^3$ km, and the duration of about 7.9 minutes. The detection of these oscillation patterns was possible thanks to the FISS capability of high temporal and spatial resolution imaging of precisely-measured Doppler velocity.  Most of the previous studies on umbral oscillations were observationally limited; some used images of intensity oscillations, but not velocity oscillations. Others worked on velocity oscillations, but the spatial coverage was limited to either one point or one dimensional array of points along the slit.

  Our results suggest that the discovered oscillation patterns are closely connected with the complex change of umbral dots. We found umbral dots beneath the oscillation centers. The umbral dots show the active motions or morphological variations and their changes are temporally associated with the duration of the oscillation patterns. With the aid of unprecedented high spatial resolution of the GST, the change of the umbral dots could be detected. As mentioned in the introduction, umbral dots themselves are regarded as the result of magnetoconvection inside umbrae. A numerical simulation shows that dark lanes related to the observed morphological changes is the results of magnetoconvective upward motion \citep{Schussler2006}. Moreover, it is natural that convective motions involve not only vertical motions, but also horizontal motions. The fast horizontal motion of umbral dots is likely to be related with the vigorous magnetoconvection. Therefore, it is reasonable to think that the rapid changes of the umbral dots are manifestations of the internal excitation events through the magnetoconvection which generate  the observed oscillation patterns.

One of the noteworthy result is that the internally excited oscillations have a peak of power in the 3 minute band.  We emphasize that because of the short period, these 3 minute oscillations in the temperature minimum region can propagate upward to appear as the umbral 3 minute oscillations observed in the chromosphere or corona. Thus we conjecture that not only the short-period tail of the external \pmode , but also the internal excitation events by magnetoconvection responsible for the umbral 3 minutes oscillations that prevail in the chromosphere.  

The oscillation patterns associated with the rapidly changing umbral dots we found strongly supports the argument from \citet{Chae2017}. They compared between spatial distribution of 3 minute oscillation power and those of umbral dots, and  argued that magnetoconvection may generate umbral 3 minute oscillations.  Even though we found only 4  oscillation patterns, we can not exclude the possibility that there may exist a more number of oscillations patterns that are too small and too weak to be clearly distinguishable. As a matter of fact, we see numerous umbral dots in both the sunspot umbrae (see Figure \ref{fig1}). Many oscillation patterns arisen from that region may be also undistinguishable if they overlap each other due to the high concentration.

The umbral oscillations generated by the internal excitation will propagate upwardly along the vertical magnetic field and develop into shock waves at the chromospheric height. They may be identified with the umbral flashes \citep{Beckers1969}. Further researches may be needed to reveal the relationship between the low atmospheric oscillation patterns originating from the internal excitations and the chromospheric umbral flashes to have a full understanding of the propagation of slow magnetoacoustic waves from the photosphere to the chromosphere in a sunspot atmosphere.

\acknowledgments
We greatly appreciate the referee's comments, which significantly contributed to the improvement of this paper. This work was supported by the National Research Foundation of Korea (NRF-2017 R1A2B4004466). 

\end{document}